\documentclass[12pt]{article}
\usepackage{epsf}

\begin{document}


\newcommand\balpha{\mbox{\boldmath $\alpha$}}
\newcommand\bbeta{\mbox{\boldmath $\beta$}}
\newcommand\bgamma{\mbox{\boldmath $\gamma$}}
\newcommand\bomega{\mbox{\boldmath $\omega$}}
\newcommand\blambda{\mbox{\boldmath $\lambda$}}
\newcommand\bmu{\mbox{\boldmath $\mu$}}
\newcommand\bphi{\mbox{\boldmath $\phi$}}
\newcommand\bzeta{\mbox{\boldmath $\zeta$}}
\newcommand\bsigma{\mbox{\boldmath $\sigma$}}
\newcommand\bepsilon{\mbox{\boldmath $\epsilon$}}

\newcommand{\vh}{{\rm v}}
\newcommand{\be}{\begin{eqnarray}}
\newcommand{\ee}{\end{eqnarray}}
\newcommand{\nn}{\nonumber}

\newcommand{\ft}[2]{{\textstyle\frac{#1}{#2}}}
\newcommand{\eqn}[1]{(\ref{#1})}
\newcommand{\vsone}{\vspace{1cm}}
 
\begin{titlepage}

\begin{flushright}
hep-th/9907014\\
KCL-TH-99-28\\
\end{flushright}
\begin{centering}
\vspace{.2in}
{\Large {\bf Dyonic Instantons in Five-Dimensional Gauge Theories}}\\
\vspace{.4in}
Neil D. Lambert and David Tong${}$ \\
\vspace{.4in}
Department of Mathematics\\
Kings College, \\
The Strand, London,\\ 
WC2R 2LS, UK\\
\vspace{.05in}
{\tt lambert,tong@mth.kcl.ac.uk}\\
\vspace{.6in}
{\bf Abstract} \\
\end{centering}
\vspace{0.2in}
We show that there exist finite energy, non-singular instanton 
solutions for five-dimensional theories with broken gauge symmetry. 
The soliton is supported against 
collapse by a non-zero electric charge. The low-energy dynamics 
of these solutions is described by motion on the ADHM moduli space 
with potential.
\vspace{.1in}


\end{titlepage}

\subsection*{Introduction}

Instantons appear as BPS particle-like configurations in five-dimensional 
supersymmetric Yang-Mills theories. If the gauge symmetry is broken 
however, the instanton shrinks to zero size resulting in a 
singular solution.  In this letter, we show that when the gauge 
group is broken to the maximal torus this collapse may be stabilised 
by a non-zero electric charge.

We consider ${\cal N}=1$ supersymmetric Yang-Mills in five 
dimensions\footnote{The ${\cal N}=1$ theory has eight supercharges. 
The arguments also apply to the ${\cal N}=2$ 
theories with sixteen supercharges providing 
four of the scalar fields are set to zero. In these 
theories, the dyonic instantons are $1/4$-BPS states.} with 
arbitrary gauge group ${\cal G}$ of rank $r$. The field 
content consists of a vector field with field strength $F$, 
a single real adjoint 
scalar $\phi$, and two pseudo-Majorana spinors. 
In five dimensions the gauge coupling constant $g$ has dimension 
$-1/2$ and the theory is weakly coupled in the infra-red. It is also 
non-renormalisable and is to be considered as an 
effective theory for some, undetermined, short-distance physics. 

The classical moduli space of the theory is ${\bf R}^r/W$, where 
$W$ is the Weyl group, and is parametrised by the 
vacuum expectation value (VEV) of $\phi$ which is taken to lie in  
the Cartan sub-algebra: 
$\langle\phi\rangle = {\bf v}\cdot{\bf H}$. We choose ${\bf v}$ 
such that ${\cal G}$ is broken to $U(1)^r$. As well as these 
surviving, local, Abelian symmetries, the theory 
contains a $SU(2)_R\times U(1)_I$ global symmetry. The  
former is the usual R-symmetry for theories with eight 
supercharges, while the latter has conserved current 
$J={}^\star{\rm Tr}(F\wedge F)$ and conserved charge equal 
to the instanton winding number, $k$. The supersymmetry 
algebra includes a single, real, scalar central charge that 
includes contributions from each local and global Abelian 
symmetry \cite{ns},
\be 
Z=\frac{8\pi^2k}{g^2}+{\bf v}\cdot{\bf q}\ ,
\label{charge}\ee
where the $r$-vector ${\bf q}$ denotes electric charge 
under $U(1)^r$. It was further shown in \cite{ns} that 
the central charge is altered by the presence of Chern-Simons terms 
which may be generated at one-loop. In this paper we shall restrict 
ourselves to actions without such terms. 

In the remainder of the paper, we shall demonstrate the existence 
of classical soliton solutions charged under both $U(1)_I$ and  
local gauge symmetries. These configurations therefore carry both 
topological and electrical charge and we christen them dyonic instantons 
in analogy with four-dimensional monopoles. 
In the simplest case of a single instanton in $SU(2)$ gauge group, 
the equations may be solved explicitly. For higher instanton charge 
in $SU(N)$ gauge group, we derive an expression for the electric 
charge in terms of the ADHM construction \cite{adhm}. We further  
describe the low-energy dynamics of instantons in spontaneously 
broken gauge groups by a 0+1 dimensional sigma-model 
on the ADHM moduli space with a potential given by the length of 
a tri-holomorphic Killing vector. The dyonic instantons appear 
as BPS solutions to the equations of motion of this massive sigma-model.

\subsection*{The Bogomol'nyi Equations}

The existence of dyonic instantons may be demonstrated by a simple 
Bogomol'nyi type argument, entirely analogous to that 
used for $1/4$-BPS states of the four-dimensional 
${\cal N}=4$ Yang-Mills theory \cite{leeyi}. One starts with the 
energy density for the bosonic fields,
\be
H&=&\ft12{\rm Tr}\ \int{\rm d}^4x\ \left\{E_\mu^2+\ft12F_{\mu\nu}^2
+({\cal D}_0\phi)^2+({\cal D}_\mu\phi)^2\right\}\ ,
\nn\ee
where $\mu=1,2,3,4$ is a spatial index, and 
$E_\mu=\partial_0 A_\mu-{\cal D}_\mu A_0$. Completing the square 
in the usual fashion, we find,
\be
H &=& \ft12{\rm Tr}\int{\rm d}^4x\left\{ 
(E_\mu-{\cal D}_\mu\phi)^2+\ft14(F_{\mu\nu}-{}^{\star}F_{\mu\nu})^2
+({\cal D}_0\phi)^2 +\ft12 F_{\mu\nu}{}^{\star}F_{\mu\nu}
+2E_\mu{\cal D}_\mu\phi\right\}\ .
\nn\ee
The final two terms are both total derivatives, an observation that 
requires use of Gauss' law,
\be 
{\cal D}_\mu E_\mu=ig[\phi,{\cal D}_0\phi]\ .
\nn\ee
If the electric field has the usual five-dimensional 
leading order behaviour for a point-like, electrically charged,  
configuration, 
\be 
E_\mu=\frac{{\bf q}\cdot{\bf H}}{2\pi^2x^4}\, x_\mu+\dots\ , 
\nn\ee
where $\dots$ denotes terms suppressed by $1/x$, then 
the resulting energy bound coincides with that derived using the 
supersymmetry algebra, $H\geq |Z|$.
The inequality is saturated by time-independent solutions 
($\partial_0=0$) with the time-component of the gauge field 
given by $A_0=-\phi$ and the remaining gauge fields 
determined by the usual self-duality condition,
\be 
F_{\mu\nu}={}^{\star}F_{\mu\nu}\ .
\label{selfdual}\ee
Finally, Gauss' law requires the adjoint scalar to obey the 
covariant Laplace equation in the background of the instanton, 
\be
{\cal D}^2\phi=0\ .
\label{d2phi}\ee
Given a self-dual field strength, there 
exists a unique solution to \eqn{d2phi} for each VEV of $\phi$, 
a result which can be seen through study of the instanton zero-modes 
\cite{leeyi}. 
In the simplest case of a single instanton in $SU(2)$ gauge group, 
solutions to both \eqn{selfdual} and \eqn{d2phi} are well 
known. In singular gauge they are given by,
\be
A_\mu=\frac{2}{g}\frac{\rho^2}{x^2(x^2+\rho^2)}\,\eta^a_{\mu\nu}x_\nu\,
\frac{\sigma^a}{2}\ \ \ \ ;\ \ \ \ \phi=\vh\frac{x^2}{x^2+\rho^2}
\frac{\sigma^3}{2}\ ,
\label{inst}\ee
where $\eta^a$ are the self-dual 't Hooft matrices, $\sigma^a$ the 
Pauli matrices in the role of $SU(2)$ gauge generators, and 
$\rho$ a collective coordinate determining the scale size of the instanton.
The five-dimensional field strength is therefore,
\be
E_\mu={\cal D}_\mu\phi = \frac{2\vh\rho^2 x_\nu}{(x^2+\rho^2)^2}
\,\eta^3_{\nu\lambda}\eta^a_{\mu\lambda}\,\frac{\sigma^a}{2}\ .
\nn\ee
Notice that in this gauge, the leading order term does not lie fully 
in the Cartan sub-algebra. The electric charge is given by the 
gauge invariant integral over the spatial boundary, 
\be
q=\int{\rm d}^4S_\mu\ \frac{1}{\vh}{\rm Tr}(\phi E_\mu)=4\pi^2\rho^2\vh\ ,
\label{q1}\ee
and a dyonic instanton of given electric 
charge $q$ is therefore seen to stabilise at size 
$\rho\sim (q/\vh)^{-1/2}$.

In order to demonstrate the BPS nature of these solutions, we 
consider the supersymmetry  transformations of a five-dimensional 
${\cal N}=1$ gauge theory. These are obtained from 
the dimensional reduction of the six-dimensional, chiral, $(1,0)$ 
supersymmetric
gauge theory. In this way we find that, for a bosonic configuration, 
supersymmetry is preserved if and only if
\be
{1\over 4}F_{\mu\nu}\Gamma_{\mu\nu}\epsilon 
-{1\over 2}\Gamma_\mu(E_\mu\Gamma_0 - {\cal D}_\mu\phi\Gamma_5)\epsilon=0\ ,
\label{BPS}\ee
where $\Gamma_0,\Gamma_1,\Gamma_2,...,\Gamma_5$ are an 
eight dimensional representation of the
Clifford algebra in six dimensions. 
The vanishing of these two terms 
therefore requires $\Gamma_{1234}\epsilon = \epsilon$ and 
$\Gamma_{05}\epsilon=-\epsilon$. Since the six-dimensional theory
is chiral (i.e. $\Gamma_{012345}\epsilon = -\epsilon$) the second condition
follows from the first and one half of the ${\cal N}=1$ supersymmetries are
thus preserved. Note that if we instead considered the maximal 
${\cal N}=2$ theory, obtained by dimensional reduction of 
the non-chiral, six-dimensional $(1,1)$ gauge
theory, then these two conditions are independent and the solution
preserves one quarter of the supersymmetry. This mirrors the 
situation for ``string-web'' type states in four-dimensions; they 
too preserve four supercharges, regardless of the number of 
supersymmetries of their parent theory.

\subsection*{The ADHM Construction}

For higher instanton winding number, explicit solutions are not 
known. For classical gauge groups however, a formal expression 
for solutions to \eqn{selfdual} may be derived using the 
ADHM construction \cite{adhm}. In recent years, this technique 
has been employed extensively by Dorey, Khoze, Mattis and collaborators 
in order to calculate instanton contributions to correlation functions 
in four-dimensional supersymmetric gauge theories. See for 
example \cite{dkm1,dkm2,kms}. Moreover, these authors have 
succeeded in casting solutions to equation \eqn{d2phi} in the language of the 
ADHM variables. We now briefly review these solutions for instantons 
of winding number $k$ in $SU(N)$ gauge groups and determine 
an expression for the electric charge of the dyonic instanton in 
terms of its collective coordinates. 

A detailed discussion of the 
ADHM construction for $SU(N)$ gauge group, as 
well as its application to solving further equations,  
may be found in \cite{kms}. The construction 
starts with a complex $N+2k\times 2k$ matrix $\Delta$. This matrix 
is linear in the spatial coordinates $x^\mu$ and, in a certain basis, 
has components, 
\be
\Delta=\left(\begin{array}{c}
\omega^a_{\ i\alpha} \\ a^{j\beta}_{\ \ i\alpha}+
x^\beta_{\ \alpha}\delta^j_{\ i}\end{array}\right)\ ,
\nn\ee
where the ranges of the indices are $a=1,\dots ,N$, $i,j=1,\dots ,k$, 
and $\alpha,\beta =1,2$. The $2\times 2$ matrix $x$ encodes the 
spatial coordinates in the usual quaternionic form, 
$x=x^\mu\sigma^\mu$, with $\sigma^\mu=(1,\sigma^a)$. The $N\times 2k$ 
matrix $\omega$ and the $2k$-dimensional square matrix $a$ are 
both independent of $x^\mu$. 

The next step is to define a matrix of orthonormal 
null vectors of $\Delta^\dagger$. This $N+2k\times N$ 
dimensional complex matrix $U$ satisfies 
${\Delta}^\dagger U=U^{\dagger}\Delta =0$. 
The ADHM ansatz for the gauge potential is then  
$A_\mu=U^\dagger\partial_\mu U$ and is seen to yield a self-dual 
field strength provided the ADHM constraints
$\Delta^\dagger\Delta=f^{-1}\otimes \sigma^0$ 
are satisfied, with $f$ a real, arbitrary, invertible $k\times k$ matrix. 
This construction can be shown to realise the full $4kN$ dimensional 
hyperK\"ahler moduli space of instantons.

In order to exhibit solutions to \eqn{d2phi} in this language, we 
first introduce a basis for the Cartan sub-algebra in which  
$\langle\phi\rangle ={\rm diag}\,(\vh^1,\dots,\vh^N)$ 
with $\sum_a\vh^a=0$. Then solutions to the adjoint Higgs equation 
of motion involves the $N+2k\times N+2k$ real, constant matrix 
${\cal A}$ and is given by \cite{dkm1, kms},
\be
\phi=U^\dagger {\cal A}U\ \ \ \ {\rm with}\ \ \ \ 
{\cal A}=\left(\begin{array}{cc}
\langle\phi\rangle^a_{\ b} & 0 \\ 0 & 
{\cal A}^{\prime\,i}_{\ \ j}\end{array}\right)\ ,
\nn\ee
where ${\cal A}^\prime$ is a real, constant $k\times k$ matrix required 
to satisfy the linear, algebraic equation
\be
\ft12{\rm tr}_2\{{\cal A}^\prime ,\omega^\dagger\omega\}
-\ft12{\rm tr}_2\left([a^\dagger,{\cal A}^\prime]a
-a^{\dagger}[a ,{\cal A}^{\prime}]\right)
={\rm tr}_2\,\omega^{\dagger}\langle\phi\rangle\omega\ ,
\nn\ee
where ${\rm tr}_2$ denotes the trace over the $\alpha,\beta=1,2$ 
indices of $\omega$ and $a$. The expression for the electric field is now 
easily determined \cite{dkm1},
\be
E_\mu={\cal D}_\mu\phi=-U^\dagger(\partial_\mu\Delta)
f\Delta^\dagger{\cal A}U-U^\dagger{\cal A}\Delta f
(\partial_\mu\Delta^\dagger)U\ .
\nn\ee
Once again employing the notation ${\bf q}=(q^1,\dots ,q^N)$, 
with $\sum_aq^a=0$, the electric charge may be expressed as 
\be
q^a=8\pi^2\sum_{\alpha=1}^{2}\sum_{i,j=1}^{k}
\left[\omega^{\dagger i\alpha}_{\ \ \ a}\omega^{a}_{\ \alpha j}
(\vh^a\delta_{\ i}^j -{\cal A}^{\prime j}_{\ \ i})\right]\ .
\label{q2}\ee
where there is no summation over the $a$ index on the right-hand 
side.

\subsection*{Low-Energy Dynamics}

The expressions \eqn{q1} and \eqn{q2} that we have derived for 
the electric charge are not quantised. This of course is to 
be expected for Noether charges in a classical theory. 
In order to see how its quantisation arises, we examine the 
low-energy dynamics of the soliton system. Again, the situation  
here is entirely analogous to the corresponding problem for 
$1/4$-BPS states \cite{blly}.

We first consider the low-energy dynamics of instantons in the 
case of vanishing  VEV, for which the full 
ADHM moduli space of solitons exists. Let this space be parametrised 
by coordinates $X^p$, (for example, $p=1,\dots ,4kN$ 
for gauge group $SU(N)$), with the corresponding instanton 
zero mode denoted as $\delta_pA_\mu$. These zero modes are subject 
to Gauss' law and must satisfy ${\cal D}_\mu\delta_pA_\mu=0$. 
The small-velocity 
interactions of instantons are described in the usual Manton 
approximation by a 0+1 dimensional sigma-model on this space with metric,
\be
g_{pq}={\rm Tr}\int{\rm d}^4x\ \left\{\delta_pA_\mu\,\delta_qA_\mu
\right\}\ .
\label{metric}\ee
This metric is singular at the points of vanishing instanton 
size. Turning now to the situation with non-zero VEV, it  
is natural to describe the low-energy dynamics in terms of the 
same sigma-model, extended to include a potential term \cite{blly},
\be
V=\ft12{\rm Tr}\int{\rm d}^4x\ ({\cal D}_\mu\phi)^2=\ft12{\bf v}
\cdot{\bf q}\ .
\label{morepot}\ee
This potential has a nice geometrical interpretation in terms 
of the ADHM moduli space metric: it is equal to half the norm 
squared of 
a tri-holomorphic Killing vector generated by the gauge transformation 
that asymptotes to ${\bf v}\cdot{\bf H}$ \cite{me}. Moreover, denoting as 
$\bepsilon\cdot{\bf K}^p$ the components of the Killing vector 
on the ADHM moduli space that is generated by the gauge transformation 
that asymptotes to $\bepsilon\cdot{\bf H}$, it was further shown in 
\cite{me} that the electric charge is given by,
\be
{\bf q}=g_{pq}({\bf v}\cdot{\bf K}^p)\,{\bf K}^q\ .
\label{q3}\ee
To see how this works in practice, we may return to the 
simplest example of the single instanton in $SU(2)$. 
In this case it is known that the instanton moduli space is 
${\bf C}^2\times{\bf  C}^2/Z_2$, where the first factor parametrises the 
centre of mass of the instanton, the radial coordinate 
of the second factor parametrises the scale size of the instanton,  
and the angular coordinates parametrise the three gauge 
orientation modes. We choose to write the 
flat metric on ${\bf C}^2/Z_2$ as the singular limit of the Eguchi-Hanson 
metric,  
\be
{\rm d}s^2 = H(r){\rm d}{\bf r}\cdot{\rm d}{\bf r}
+H^{-1}({\rm d}\psi +\bomega\cdot{\rm d}{\bf r})^2\ ,
\nn\ee
where $H=2/r$ and $\nabla\times\bomega =\nabla H$. Furthermore, 
$\psi$ has range $0\leq\psi <4\pi$ and $r=\pi^2\rho^2$, where       
the coefficient $\pi^2$ has been determined in 
order to derive the correct numerical value for the potential below. 
The singularity of the metric at $r=0$ reflects the singularity of 
the corresponding instanton solution \eqn{inst} at $\rho =0$. An 
advantage of the above description of ${\bf C}^2/Z_2$ is 
that the tri-holomorphic Killing vector that arises 
from $U(1)$ gauge transformations (with period $2\pi$) 
is manifest and is given by $2\partial_\psi$. The potential 
\eqn{morepot} becomes, 
\be
V=4\vh^2g_{\psi\psi}=4\vh^2 H^{-1} =  2\pi^2 \vh^2\rho^2\ ,
\label{pot}\ee
whose functional form is in agreement with \eqn{q1} and \eqn{morepot}. 
Notice that, unlike the similar 
potential on Taub-NUT space that appears in the low-energy 
effective dynamics of $1/4$-BPS states \cite{me, blly}, the 
instanton potential \eqn{pot} is unbounded.

So far, we have only described the low-energy dynamics of 
instantons embedded in broken gauge groups. We have yet to 
recover a description of the dyonic instantons. In fact, these 
appear as solutions to the equations of motion of the massive 
sigma model \cite{blly,dt}. Indeed, one may 
easily verify that solutions to the first order equation,
\be
\frac{{\rm d} X^p}{{\rm d} t}={\bf v}\cdot{\bf K}^p\ ,
\label{sigbog}\ee
also satisfy the equations of motion. Compactness of the 
gauge group ensures that the orbits 
of these solutions are compact and they therefore have 
the interpretation of excited instanton states rather than 
scattering states. They have mass given by $H={\bf v}\cdot{\bf q}$ 
and are identified with the dyonic instantons.  

Let us now consider the fermions in our discussion. 
Their properties follow from supersymmetry and the BPS nature \eqn{BPS} 
of the dyonic instantons. The bosonic collective coordinates 
are accompanied by $4kN$ fermionic collective coordinates 
(this number is increased to $8kN$ for the ${\cal N}=2$ theory) 
and the massive sigma model described above is supersymmetrised 
accordingly. Indeed, a potential 
given by the length of a tri-holomorphic Killing vector, 
as in equations \eqn{morepot} and \eqn{q3}, is of the form that 
allows a supersymmetric completion of up to 8 supercharges \cite{ag} 
and it is easily seen that solutions to \eqn{sigbog} 
saturate the BPS-bound of the sigma-model supersymmetry algebra, 
reflecting their BPS nature in the underlying five-dimensional 
theory. In terms of the ADHM language, expressions for the 
(non-kinetic) fermion terms of the massive sigma model, as 
well as the supersymmetry transformations, have been constructed 
in \cite{dkm1,dkm2,kms}.

For the ${\cal N}=1$ theory, one may consider the addition of 
hypermultiplets in various representations of the gauge group. 
There exists a single 
real, bare mass parameter, $m$, which may be assigned to each 
hypermultiplet. Restricting ourselves to $N_f$ hypermultiplets 
in the fundamental representation of an $SU(N)$ gauge group,  and 
denoting the bare masses in $N_f$-vector form $\vec{m}$,  
these parameters generically break the 
global flavour group to its Cartan sub-algebra and induce a 
further term in the central charge of the form 
$\vec{m}\cdot\vec{S}$ 
where $\vec{S}$ denotes the charge of a state under the surviving 
Abelian flavour symmetries. An 
examination of the instanton mass \cite{dkm2,kms} reveals that 
the instanton is indeed charged under these flavour symmetries, 
courtesy of fundamental fermion zero modes.

We may now turn to the question of the BPS spectrum in the 
quantum theory. In the case of vanishing VEV, naive semi-classical 
quantisation of the scaling collective coordinate would  appear to 
lead to a continuous spectrum of particles. The interpretation of this 
is not well understood. With non-vanishing VEV, the problem is different. 
Uncharged instantons become singular and their existence 
in the theory can only be determined by appealing to new 
ultra-violet degrees of freedom, generically arising from string theory. 
This reflects the non-renormalisability of five-dimensional Yang-Mills theory. 
Dyonic instantons 
however suffer from neither of these problems and it is  
sensible to attempt to determine their spectrum without recourse to 
string theory. Quantisation of sigma-models with potential has 
been considered recently in \cite{blly} in the context of $1/4$-BPS states. 
These authors find that the states of the theory are described by 
suitable, normalisable differential forms satisfying,
\be
i(d - i_{{\bf v}\cdot{\bf K}})\Omega = 
\pm\star(d - i_{{\bf v}\cdot{\bf K}})\star\Omega\ ,
\nn
\ee
where $i_{{\bf v}\cdot{\bf K}}$ denotes contraction with the 
Killing vector field ${\bf v}\cdot{\bf K}$ and $\star$ is the 
Hodge dual. Moreover, the 
quantisation of electric charge now becomes apparent arising, 
as for the more familiar dyons in four-dimensions, through the 
quantisation of a periodic variable parametrising the orbits of 
the Killing vector fields. 

The BPS spectrum 
of five-dimensional gauge theories has been derived in the 
context of fivebrane webs (see for example \cite{ahk}) where 
the minimal instanton states were indeed found to be dyonic. 
It would be interesting to reproduce these results from 
field theory. 

Finally, we turn to six-dimensional theories with sixteen supercharges. 
Consider first 
the $(1,1)$ theory that propagates on the IIB fivebrane. The methods 
in this paper may be employed to construct an instanton string in this 
theory. 
The identification of this soliton 
with a bound state of D-strings and fundamental strings will 
be explored in a forthcoming publication \cite{lt}. 
The relationship to the $(2,0)$ theory that exists on coincident 
IIA fivebranes is more subtle: the 0+1-dimensional ADHM  sigma model 
with eight supercharges has 
been proposed as a matrix model description of this theory \cite{abkss}. 
The massive sigma model described 
above corresponds to the spontaneously broken theory on separated 
fivebranes. It is natural to interpret the states described by \eqn{sigbog} as
``W-boson'' strings of the 2-form gauge 
potential. 

\subsection*{Acknowledgements}

D.T. is supported by an EPSRC fellowship. We are especially 
grateful to 
Bobby Acharya and Jerome Gauntlett for conversations and  
Dongsu Bak, Choonkyu Lee, Kimyeong Lee and Piljin Yi for 
correspondence.

\end{document}